\documentclass[12pt]{article}
\usepackage[table,xcdraw]{xcolor}
\usepackage{blindtext}
\usepackage[utf8]{inputenc}
\usepackage{graphicx}
\graphicspath{ {images/} }
\usepackage{gensymb}
\usepackage{fancyhdr}
\usepackage{placeins}
\usepackage{amsmath}
\usepackage{tcolorbox}
\usepackage{rotating}
\usepackage{xltabular}
\usepackage{booktabs}
\usepackage[numbers]{natbib}

\usepackage[english]{babel}

\usepackage[letterpaper,top=2cm,bottom=2cm,left=2cm,right=2cm,marginparwidth=2.00cm]{geometry}

\usepackage{amsmath}
\usepackage{graphicx}
\usepackage[colorlinks=true, allcolors=blue]{hyperref}

\newcommand{\cmmnt}[1]{}

\title{Computational optimisation of slow cooling profiles for the cryopreservation of cells in suspension}
\author{Jack Lee Jennings$^1$, Sanja Bojic$^2$, Lukas Breitwieser$^{3, 4}$, Alex Sharpe$^5$, Roman Bauer$^6*$}

\date{\footnotesize Interdisciplinary Computing and Complex Bio-Systems, School of Computing, Newcastle University, Newcastle Upon Tyne, UK $^1$.\\
Stanford University School of Medicine, Stanford, USA $^2$.\\
IT Department, CERN, Geneva 1211, Switzerland $^3$.\\
Department of Computer Science, ETH Zurich, Zurich 8092, Switzerland $^4$.\\
Biosciences Institute, Faculty of Medical Sciences, Newcastle University, UK $^5$.\\
Nature Inspired Computing and Engineering (NICE) Research Group, Computer Science Research Centre, University of Surrey, Guildford, UK $^6$.\\
* Corresponding author.
}

\begin{document}
\maketitle

\begin{abstract}
The cryopreservation of biological materials is a highly complex process, as it involves numerous factors such as the cooling and thawing procedures, the administration of cryoprotective agents (CPAs), as well as the type and composition of cells. While theoretical work has yielded a better understanding of the processes occurring during cryopreservation, the design of cryopreservation protocols and their parameters is currently predominantly based on heuristic optimization. Here, we propose a mathematical method to optimise the cooling dynamics in slow-cooling, to reduce the risk of injury. We derive our method from first principles and provide computational predictions. Moreover, we assess the predictions with data obtained from the literature, as well as novel experimental results. Overall, we provide a generic computational approach to generate improved slow-cooling profiles for the cryopreservation of cells in suspension.

\end{abstract}
\section{Introduction}

The cryopreservation of cells is currently used in numerous biomedical applications \cite{bojic2021winter}. However, significant challenges remain and constitute a bottleneck. A key use case for cryopreservaiton is its possible applications in the successful long term storage of tissue and cells. Representing an exciting prospect for biomedical research as a whole\cite{bojic2021winter}. Stem cells, for example, have great potential for research due to their ability to be differentiated into a wide variety of cells and tissue but cannot be stored in a standard manner for sufficient time lengths with consistent quality and quantity \cite{cryopause}.  The banking of such cells would allow for a ready supply of tissue for use in pharmaceutical research rather than requiring fresh sample equivalents \cite{celltherapy}. Toxicology testing of newly designed drugs being a prime example of a pharmaceutical process requiring a readily available source of tissue analogues \cite{toxicol}. Additionally,  the availability of suitable biological material for testing would reduce the need for animal based counterparts.     

The storage of cells used for in vitro fertilisation represent another key area of interest for improved preservation protocols \cite{reproduct}. Outside of the obvious human need, the preservation of genetic material of endangered species will be of key importance in the future for use in preservation efforts \cite{endangered}. However, the differential storage techniques required for each individual species cells makes this difficult, with some species reproductive cells still unable to be preserved reliably. This is due to different cell and tissues types requiring various combinations of cooling rates, cryoprotective agents (CPA) concentrations and thawing trajectories. This renders the design of cell-type specific experimental protocols an endeavour that is experimentally expensive and time-consuming. 

The two best-established methods for cryopreservation are vitrification and slow-cooling. Vitrification is a relatively new method where high cooling rates are employed. To this end, usually liquid nitrogen is used. However, vitrification requires high concentrations of cryoprotective agents (CPAs) to inhibit intracellular and extracellular ice formation (IIF and EIF). Vitrification can result in excellent post-thaw survival rates (80 - 95.6\%) \cite{loutradi2008cryopreservation,amps2010situ}. However, vitrification is problematic with regard to medical applications due to lack of scalability, high CPA concentration requirements and possibility of sample contamination through direct contact with liquid nitrogen \cite{STEM,vajta2006programmable,consuegra2019vitrification}. 

Slow-cooling cryopreservation protocols avoid IIF through the use of tailored combinations of cooling rates and CPA concentrations. Notably, recent work shows that slow-cooling can provide comparable post-thaw survival results for ovarian tissue and other cryopreserved biological materials to that of vitrification and state of the art methods \cite{lee2019comparison,kulikova2019survivability}. A major issue with both methods however is the need for extensive trial-and-error based wet-lab work to optimise slow-cooling protocols for every cell type.

Some studies have shown, when using slow cooling rate techniques, employing stepped accelerating cooling profiles, can achieve greater post-thaw survival results for spermatozoa than standard liquid nitrogen plunge cooling profiles \cite{galarza2019two}. Indicating that using an initially low cooling rate into a secondary faster cooling rates could provide another avenue for avoiding intracellular ice formation in cells undergoing cryopreservation. As cells are given enough time to remove large proportions of their intracellular water content before ice nucleation, thus lethal intracellular ice formation can be avoided. This multi-step approach also allows for cells avoid death due to high extracellular concentrations. As higher cooling rates at the later accelerating stage lead to rapid cellular cooling possibly avoiding possible negative effects due to large exposure time to high solute and cpa concentrations. This could be critical to improving the post-thaw survival of some cell types such as mouse spermatozoa which have previously been shown to have poor post-thaw survival results when using classical slow cooling techniques \cite{koshimoto2002effects}. Currently however, there is very little work in regards to the testing of such multi-step cooling rates or how it may impact different cell types. This is due to the difficulty in optimising such cooling profiles experimentally, as the addition of one additional cooling rate necessitates testing across a much wider range of variables than single step. The addition of even further cooling rates further compounding this issue to the point of being un-viable to test physically.

Computational modelling of the cryopreservation process represents an excellent solution to this optimisation problem. As a computational model has the capability to simulate a vast number of cooling rate and switching temperature combinations in a relatively short time scale. In addition, the use of a computational model in this effect would allow for the investigation and interrogation of the relationship between cells and stepped cooling profiles that would be unfeasable for experimental work. Previous work has already shown that computational models may be able to predict multi-stepped cooling profiles that lead to improved post-thaw survival of cryopreserved cells\cite{STEM}. However, computationally estimated stepped cooling profiles have not been validated experimentally. 

Two techniques have been shown to yield satisfactory results for the simulation of cellular responses during cooling. The first is a hybrid mathematical model used for simulating the cellular response to cryopreservation using physical cellular properties, temperature conditions, ice formation and chemical diffusion. Variations of this model have already been used  been used  to simulate mass and heat transfer in cryopreserved cells \cite{skorupa2020numerical} and ice formation within cells \cite{cincotti2012modelling, GYU}. This type of model was recently expanded upon and used by Hayashi \textit{et al} to predict the post-thaw survival of cryopreserved human induced pluripotent stem cells (hIPSCs) \cite{STEM}. However, currently Hayashi \textit{et al}'s variant of the model is still missing two important things. Firstly, they do not compare the ability of their model for predicting the post-thaw survival of cells against multiple cell types. Secondly, Hayashi \textit{et al}'s work despite predicting an optimal cooling procedure does not validate the cooling profile experimentally. 

The second technique uses a stochastic Differential Evolution algorithm (DEA), as defined by Storn and Price \cite{storn1997differential}. This method combines computational modelling with iterative experimental feedback to optimise cooling rates and CPA concentrations through an evolving algorithm. DEAs have shown success in the work of Tsutsui \textit{et al} and Pollock\textit{et al} for predicting and optimising the post-thaw survival of multiple cell types \cite{tsutsui2011optimized,pollock2017algorithm}. However, a major issue with the DEA method is its need for repeat experimental feedback in an iterative cycle between experimental and computational data. Thus, it reduces one of the major benefits of optimising protocols computationally, i.e. the significant reduction in costs related to repeated experimental data acquisition. 

Thus, we here present a computational model for simulating the cryopreservation of cells in suspension with a focus on optimising experimental procedures \textit{in silico} \cite{jennings2024virtual}. Necessary parameters for cell properties are taken from relevant literature with comparisons made against pre-existing experimental data. Ultimately, we have utilised our model to investigate three major questions: (1) Can cryopreservation simulations be used to predict the post-thaw survival of multiple cell types? (2) Can computationally optimised cooling profiles improve experimental post-thaw survival (3) Do multi-stepped cryopreservation cooling protocols predict for better survival results than standard single cooling rate procedures?         

\section{Methodology}
\subsection{Computational modelling and simulation}

\begin{figure}[h]
	\centering
	\includegraphics[scale=0.5]{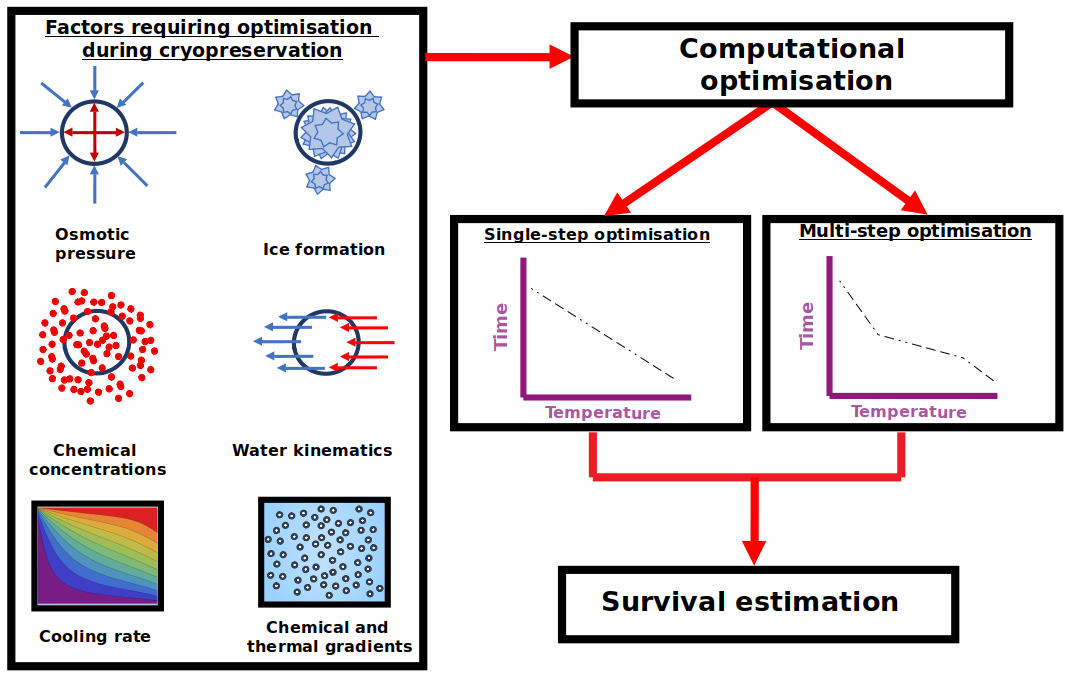}
	\caption[A total of at least 7 major factors must be considered for the simulation of cells undergoing cryopreservation as outlined above. This can then ultimately be used to either simulate and optimise a single variable for maximum post-thaw survival or the optimisation of multiple variables.]{Our computational framework utilises a hybrid model for simulating and optimising the response of cells to various cryopreservation protocols. The framework uses initial cellular and environmental variables to simulate the cryopreservation of cells and then predict post-thaw survival chance for each individual cell in the suspension. The results from our algorithm can be used to optimise the cryopreservation protocols used for both standard single stepped and multi-stepped cooling approaches.}
	\label{Simple model}
\end{figure}

We utilise a multi-factor mechanistic model for predicting the post-thaw survival of cells in suspension undergoing cryopreservation, as first hypothesised by P. Mazur \cite{mazur1960}. Our model for the response of cells in suspension to various cryopreservation procedures takes a hybrid approach. An outline of how our algorithm works can be found in Figure \ref{Simple model}.

\subsubsection{Mass transfer model}
Our model for the transport of water is based upon a simplified version of P. Mazur's (1963) water transport equation derived by G. M. Fahy (1981)\cite{FAHY1981473}.
\\
\begin{equation}
\centering
\frac{dV}{dT}=\frac{L_pART}{B v^0_1}\left [ \frac{\Delta H_f}{R}\left ( \frac{1}{T_0} - \frac{1}{T} \right ) - ln\left ( \frac{V}{V + n_sv_w} \right )  \right ]
\label{Volume equation}
\end{equation}
\\
Here V is cellular water volume ($\mu m^3$), $v_w$ is the partial molar volume of water, t is time (minutes), $L_p$ is the hydraulic permeability of the cellular membrane ($\frac{\mu m}{min \cdot atm}$), $A$ is the effective surface area of the cell ($um^2$), R is the universal gas constant, T is the temperature (K), $\Delta H_f$ is the latent heat of fusion of water and $n_s$ is moles of solute inside the cell and the molar volume of water $v^0_1$. See also Suppl. Table S1 for clarifications on variables and abbreviations used in this work.
The hydraulic permeability of cells is not constant, we use the Arrhenius relation to temperature as follows:
\\
\begin{equation}
    \centering
    L_p = L_{p0}exp{\frac{-E_a}{R}\left [ \frac{1}{T}-\frac{1}{T_0}\right ]}
    \label{Permeability equation}
\end{equation}

 The transport of permeating CPA's into intracellular space follows a similar transport equation based on the one flux equation of Whittingham \textit{et al} (1972) \cite{Whittingham411} with simplifications made by Fahy \textit{et al} (1981) \cite{FAHY1981473}.

\begin{equation}
\centering
\frac{dn_e}{dt}=p_{s}A n_{s}\times 10^{15} \left ( \frac{1}{V_{eq}} - \frac{1}{V}\right )
\label{Solute equation}
\end{equation}

Here n is moles of solute, $p_{s}$ is the cell membrane permeability coefficient to the given CPA and $V_{eq}$ is the equilibrium volume of the cell. $p_{s}$ like $L_{p}$ is also temperature dependant and varies as follows:

\begin{equation}
    \centering
    p_s = p_{s_0}exp{\frac{-E_{as}}{R}\left [ \frac{1}{T}-\frac{1}{T_0}\right ]}
    \label{Solute permeability}
\end{equation}
\\
 
\subsubsection{Heat transfer and chemical concentrations}

For the modelling of heat transfer and chemical concentration we utilise the frameworks for 3D diffusion environments provided by the BioDynaMo computational framework (BioDynaMo v1.01.115-e1088d4a)\cite{breitwieser2022biodynamo,breitwieser_biodynamo_2023}.

This form of diffusion and heat transfer is based on the 1D heat transfer equation:

\begin{equation}
\centering
\frac{\delta u}{\delta t} = \alpha \frac{\delta^2 u}{\delta x^2}
\label{1D Chemical}
\end{equation}

 Chemical concentration is represented by u, t is time, $\alpha$ is the diffusion coefficient and x is our spatial parameter. Equation \ref{1D Chemical} can then be mapped onto a grid of finite difference as follows through time:

\begin{equation}
\centering
\left ( \frac{\delta u}{\delta t} \right )^{n}_{i} \approx \frac{u^{n+1}_i - u^{n}_i}{\Delta t}
\label{Time 1}
\end{equation}

\begin{equation}
\centering
\left ( \frac{\delta ^2u}{\delta x^2} \right )^{n}_{i} \approx \frac{u^{n}_{i+1} - 2u^{n}_i +u^{n}_{i-1}}{\Delta x^2}
\label{Time 2}
\end{equation}

\begin{equation}
\centering
u^{n+1}_i = u^{n}_i+ Q\left ( u^{n}_{i+1} -2u^{n}_{i} + u^{n}_{i-1} \right ) 
\label{Time 3}
\end{equation}

 Q represents the Courant-Friedrichs-Lewy (CFL) number.
 
 \begin{equation}
\centering
Q = \frac{\alpha \Delta t}{\Delta x^2} 
\label{CFL}
\end{equation} 

We then expand our equation to 3D space through the z and y direction, hereby creating a new space of length L in the x,y and z directions. This is further broken down into individual local spaces of side length $\Delta$x, $\Delta$y and $\Delta$z. Assuming our space is made of cubes with equal side lengths we get the 3D variation of Equation \ref{Time 3}:

\begin{equation}
\centering
\Delta x^2 = \Delta y^2 = \Delta z^2
\label{Space equation}
\end{equation}

\begin{equation*}\label{3D Diffusion}
\begin{aligned}
\centering
u^{n+1}_{i,j,k} &= u^{n}_{i,j,k}+ Q(( u^{n}_{{i+1,j,k}} -2u^{n}_{{i,j,k}} + u^{n}_{{i-1,j,k}})\\
 & + ( u^{n}_{{i,j+1,k}} -2u^{n}_{{i,j,k}} + u^{n}_{{i,j-1,k}})\\
 & +( u^{n}_{{i,j,k+1}} -2u^{n}_{{i,j,k}} + u^{n}_{{i,j,k-1}} ))
\end{aligned}
\end{equation*}

We similarly convert the CFL number to its 3D form. This is then used as our equations stability criterion for the partial differential equation as follows:

\begin{equation}
\centering
|Q| < \frac{1}{6} \:\:\:\:\:\: or \:\:\:\:\:\: \frac{\alpha \Delta t}{\Delta x^2} < \frac{1}{6}
\label{CFL 3D}
\end{equation}   

This variant of the diffusion equation is often refereed to as the Euler method. 

\FloatBarrier
\subsection{Experimental procedures}

\subsubsection{Cell cultures}
 For our wet-lab experiments, we used human T lymphoblasts - Jurkat cells (Clone E6-1; TIB-1522; American Type Culture Collection, Manassas, VA). Jurkat cells were cultured in a complete growth medium containing RPMI 1640 Medium (A1049101; Thermofisher Scientific, Massachusetts, US), 10\% fetal bovine serum (10270106; Thermofisher Scientific, Massachusetts, US), and 1\% penicillin-streptomycin (15140122; Thermofisher Scientific, Massachusetts, US) in a tissue culture incubator at 37\degree C with a humidified atmosphere containing 5\% CO2. Cells were grown in suspension in T75 tissue culture flasks and maintained at a concentration between $1\times10^{5}$ and $1\times10^{6}$ cells mL$^{-1}$, as recommended by the supplier.
\subsubsection{Freezing and thawing}
 For cryopreservation experiments, Jurkat cells were dispensed into 2mL cryotubes at $3.0\times10^{6}$ cells mL$^{-1}$, in a complete medium with the addition of 10\% DMSO (102952; Merck, Darmstadt, Germany) as a cryoprotective agent. Cells were cryopreserved using a Kryo 360 controlled rate freezer (Planer, Middlesex, UK) at various cooling rates (10\degree C $min^{-1}$, 3\degree C $min^{-1}$, 0.5\degree C $min^{-1}$) or using our three-step cooling method. All the experimental conditions were tested as both technical and biological triplicates. Before freezing, the Kryo 360 controlled rate freezer was cooled down to +4\degree C, which was used as a starting temperature for all the experiments. After reaching -50°C, cells were kept frozen for an additional one hour. Afterward, vials were quickly thawed in a 37\degree C water bath.
\subsubsection{Staining and cell counting}
 As soon as an experimental vial content was thawed, the cell suspension was transferred to a centrifuge tube containing 9 mL complete medium and centrifuged at 1000 rpm for 5 minutes. After spinning cells down, the supernatant was removed, and the remaining cell pellet was re-suspended in the complete medium, dispensed into a 75 cm$^{2}$ tissue culture flask, and incubated in a tissue culture incubator at 37\degree C with a humidified atmosphere containing 5\% CO2. The living cells were counted using a trypan blue exclusion test, 1 hour and 24 hours after thawing, with at least 200 total cells counted for each experimental condition tested.
 
 \subsubsection{Software and hardware resources}
 All simulations were written in the C++ programming language on and ran on a standard Stoner workstation with an Intel core i7 processor, 8GB of memory and a GeForce 730 graphics card. The operating system (OS) used by our system was Ubuntu 18.04. 
\FloatBarrier
\section{Results}

 \begin{figure}[h]
	\centering
	\includegraphics[scale= 0.9]{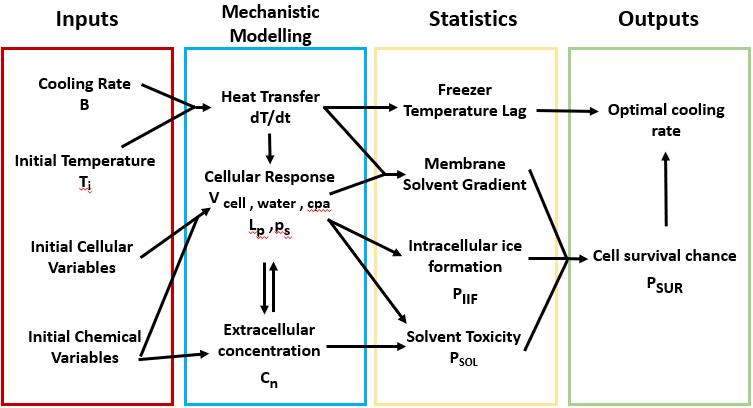}
	\caption[Chart showing modelling approach]{An overview of the hybrid modelling approach taken by our computational framework with four major sections which are handled during simulation. Inputs are all initial variables which are required for modelling such as: cell type, CPA, extracellular concentrations etc. Mechanistic modelling handles all of the advancing simulation parameters such as heat transfer, mass transfer and cellular responses. Statistics includes any calculations for probabilities and similar events. Finally, outputs are the final variables calculated during simulation and output at the final step.}
	\label{Model overview}
\end{figure}

 Here we present the outputs from our computational framework for predicting the post-thaw outcomes of multiple cells in suspension to various cryopreservation protocols. Our computational framework uses a mechanistic approach for simulating  cellular response, intracellular ice formation and extracellular solutes. Predictions made using these variables are compared to available experimental literature and our own experimental work. An outline of the computational framework can be found in Figure \ref{Model overview}. We derive a computational model for survival by combining the probabilities for both intracellular ice formation and solute injury.
 
\subsubsection{Multi-stepped cooling}

 In addition to standard cooling, we also utilise cooling profiles that include multiple steps. Standard cooling protocols of samples were simulated from an initial temperature $T_i$ to a final hold temperature $T_f$ with a singular cooling rate $CR_1$. Stepped cooling utilised additional switching temperature variables $T_{switch N}$ and more than one cooling rate. In the scenario of n-step cooling, cells are first cooled at an initial cooling rate $CR_1$ until reaching a temperature of $T_{switch 1}$ with cooling then being switched to $CR_2$. This is then stepped to the n-th level based on the number of desired cooling steps. 

Mathematically, multi-step cooling is handled in the following way to the n-th switching temperature:

 \begin{equation}
 \centering
 Cooling \:rate (B) = \left\{\begin{matrix} CR_1 & if \:\: T\geq T_{switch\:\:1} \:\: 
  & \\ CR_2 &  if \:\: T<  T_{switch\:\:1}  \:\:and\:\: T\geq  T_{switch\:\:2}\:\:
  & \\...
  & \\ CR_N & if \:\: T<  T_{switch\:\:N-1} \:\:
 \end{matrix}\right.
 \label{Multi-cooling equation}
 \end{equation} 
 
 \subsubsection{Intracellular ice formation}
Our model for intracellular ice formation is based upon the two-factor hypothesis of surface catalysed nucleation (SCN) and volume catalysed nucleation (VCN) \cite{PITT198944,MTONER,GYU}. Using the two-factor model we can express the probability of IIF occurring within any given cell as follows: 

\begin{equation}
\centering
P_{IIF} = P^{SCN}_{IIF}+\left ( 1 - P^{SCN}_{IIF} \right )P^{VCN}_{IIF}
\label{IIF Probability}
\end{equation}

Both SCN and VCN ice nucleation is defined as follows:

\begin{equation}
\centering
P^{XCN}_{IIF} = 1 - exp\left [ -\frac{1}{B} \int_{T_0}^{T}ZI_{XCN}\:dT \right ]
\label{IIF XCN}
\end{equation}

with XCN representing SCN or VCN. For SCN Z is the effective surface area of the cell whereas for VCN Z is the cellular volume. I$_{XCN} $ is the nucleation rate of ice as described in the work of J. Yi et al (2014) \cite{GYU}. ZI$_{XCN} $ represents the probability density of XCN type ice formation. Our integer portion of the equation can be expressed in terms of either time or temperature: 

\begin{equation}
\centering
 -\frac{1}{B} \int_{T_0}^{T}ZI_{XCN}\:dT = -\int_{0}^{t}ZI_{XCN}\:dt
 \label{Time to Temperature}
\end{equation}

We also apply conditional probability to our model for ice formation. This is done to account for those situations where the formation of ice is impossible within the cell. Firstly, if the cell has reached its minimum volume ($V_{cell} = V_b$) no further ice formation can occur due to lack of available intracellular water. Secondly,  if the freezing temperature of water $T_f$ has not been reached no phase change can occur. Finally, as stated by P. Mazur in 2004, cells require a minimum amount of osmotically interacting water for lethal intracellular ice formation \cite{frozenstateMazurIce}. This is due to protoplasmic macro molecules reaching extreme concentrations if less than 10 to 15\% of osmotic water remains, which ultimately prevents lethal IIF. To account for this we use the following conditional probabilities:

\begin{equation}
\centering
P_{IIF}= 
\begin{cases}
P^{SCN}_{IIF} + (1-P^{SCN}_{IIF})P^{VCN}_{IIF}     &\text{if} \:\:V_{cell} > V_b \:\:  \text{,}\:\:  T \leq T_f \:\: \text{and}\:\:  V \geq  0.1 \cdot V_i \\
0              &\text{if}\:\: V_{cell} = V_b \:\:,  \:\: T > T_f \:\:\text{or} \:\:V < 0.1 \cdot V_i \\
\end{cases}
\label{IIF conditional}
\end{equation}

\subsubsection{Toxicity and membrane damage}

Cell membrane damage was assumed to be caused by the excessive efflux of water during cooling or influx during re-warming. This leads to cellular membrane damage due to an inability to dilute extracellular concentrations through excretion of water. We evaluate this as through the rate of change of cell surface aread during cryopreservation, first quantified by G. Fahy (1981) \cite{FAHY1981473} as follows:  

\begin{equation}
\centering
I_{SOL} = {CR}\frac{\Delta\:V}{\Delta\:T}\times\frac{1}{V_0}\times\frac{1}{A}
\label{Solution I}
\end{equation}

The extracellular concentrations build up around cells during cooling due to the formation of extracellular ice pushing cells and solvents/solutes closer together. Cells that are exposed to these high concentrations for extended periods suffer toxic damage to their cellular membrane. This can be seen as a cell having a surface area that is too low compared to its initial area, remaining too close to the equilibrium volume for extended periods of time. This ultimately leads to an inability to suitably excrete water into the extracellular environment and dilute surrounding solutes. 

From here we can begin to generate a probability function for a cell experiencing a toxic level of solution buildup at the cell membrane boundary.

\begin{equation}
\centering
P_{SOL} = 1 - exp\left [ -\frac{1}{B} \int_{T_0}^{T}AI_{SOL}\:dT \right ]
\label{Solution probability}
\end{equation}

\subsubsection{Survival predictions}

For our computational framework we have defined the chance that a cell dies during the cryopreservation process. To do this we first defined a general probability for a cell surviving as follows: This can be seen as the certainty of a cell surviving (or 1) minus the probability of a lethal effect occurring during cryopreservation:

\begin{equation}
	\centering
	P_{SUR} = 1 - P^{Lethal}
	\label{Survival probability simple}
\end{equation}

$P^{Lethal}$ here represents the general combination of all probabilities which lead to cell death during cryopreservation. Within our computational framework we define $P^{Lethal}$ as a combination of lethal IIF ($P^{Lethal}_{IIF}$) and/or lethal chemical injury ($P^{Lethal}_{SOL}$) occurring. Thus we can define $P^{Lethal}$ as follows:

\begin{equation}
\centering
P^{Lethal} = P^{Lethal}_{IIF} + P^{Lethal}_{SOL} - P^{Lethal}_{IIF} \times P^{Lethal}_{SOL}
\label{Lethal probability}
\end{equation} 

This can be understood as the sum of the probabilities for lethal IIF and lethal chemical injury occurring minus the chance of both effects occurring at the same time. Using Equation \ref{Survival probability simple} and \ref{Lethal probability} we can therefore define the probability of a single cell surviving the cryopreservation process as follows:

\begin{equation}
\centering
P_{SUR} = 1 - (P^{Lethal}_{IIF} + P^{Lethal}_{SOL} - P^{Lethal}_{IIF} \times P^{Lethal}_{SOL})
\label{Survival probability expanded}
\end{equation}

Using Equation \ref{Survival probability expanded}, we can therefore predict the chance any individual cell within our simulation dying at any given time step.

\subsection{Standard cooling modelling}
\begin{figure}[h]
	\centering
	\includegraphics[scale=0.55]{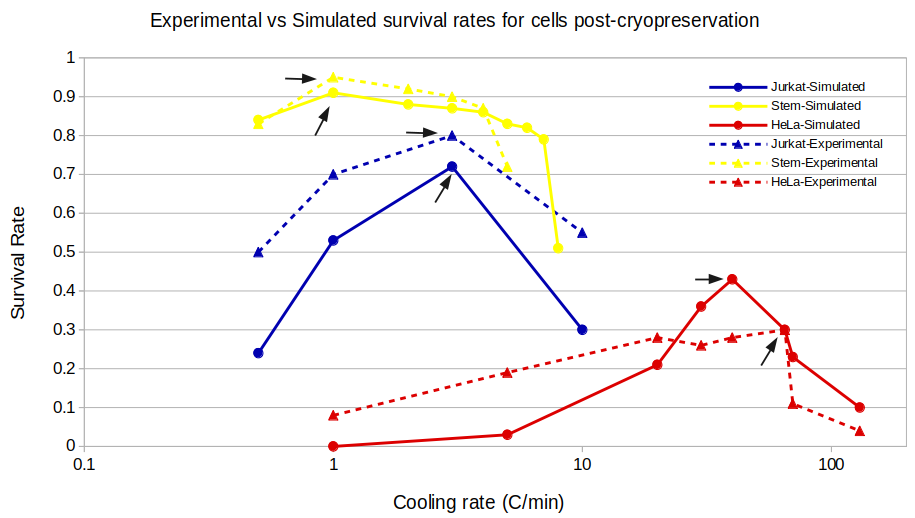}
	\caption[Triple survival curves]{ Post-thaw survival rate curves for Jurkat (Blue), HeLa cells (Red) and Human induced pluripotent stem cells (Yellow). Experimental (dashed) and simulated (solid) lines. Arrows indicate optimum cooling rates for post-thaw survival of cells.}
	\label{Simulated cell survivals}
\end{figure}

 For standard cooling, we predict the post-thaw survival of three cell types: Jurkat, HeLa and human induced pluripotent stem cells (HiPCs). These results are then compared against the experimental results of Yu \textit{et al} \cite{YU20172653} for Jurkat cells, Hayashi \textit{et al} \cite{STEM} for stem cells, and McGrath \textit{et al} for HeLa cells \cite{mcgrath1975experimental}. Simulations for the cryopreservation of each cell type followed the same protocols as stated in their respective works. Jurkat simulations were performed with cooling rates from 0.1 to 10 \degree C$min^{-1}$. HeLa simulations were performed with cooling rates from 1 to 120 \degree C$min^{-1}$. Finally, HiPSCs were simulated with cooling rates from 0.1 to 10 \degree C$min^{-1}$. 
 The predictions for the post-thaw survival of cells undergoing standard freeze/thaw procedures compared against experimental literature are displayed in Figure \ref{Simulated cell survivals}. Maximum post-thaw survival for simulated and experimental results are highlighted by black arrows.

\FloatBarrier
\subsection{Multi-step cooling modelling}

 In addition to standard linear cooling of Jurkat cells we also used our computational framework to investigate two- and three-stepped cooling. As previously, simulations were performed with 10\% DMSO at cooling rates ranging between 0.1 to 10 \degree C$min^{-1}$. Multi-step cooling simulations including an additional switching temperature parameter between -0.1 to -40 \degree C. For each switching temperature a unique relationship was found between the switching temperatures and the final predicted survival of cells.
 
 Here, we utilised a grid search technique for parameter optimisation. To this end, a search across switching temperatures and cooling rates was implemented to find the best cooling profile for maximising post-thaw survival rates. For two- and three-step cooling a total of 4,000 and 1.6 million simulations were performed respectively. Hence, a total of 2 million simulations were performed for the cryopreservation of Jurkat cells. 
 
\begin{sidewaysfigure}
    \centering
    \includegraphics[scale= 0.70]{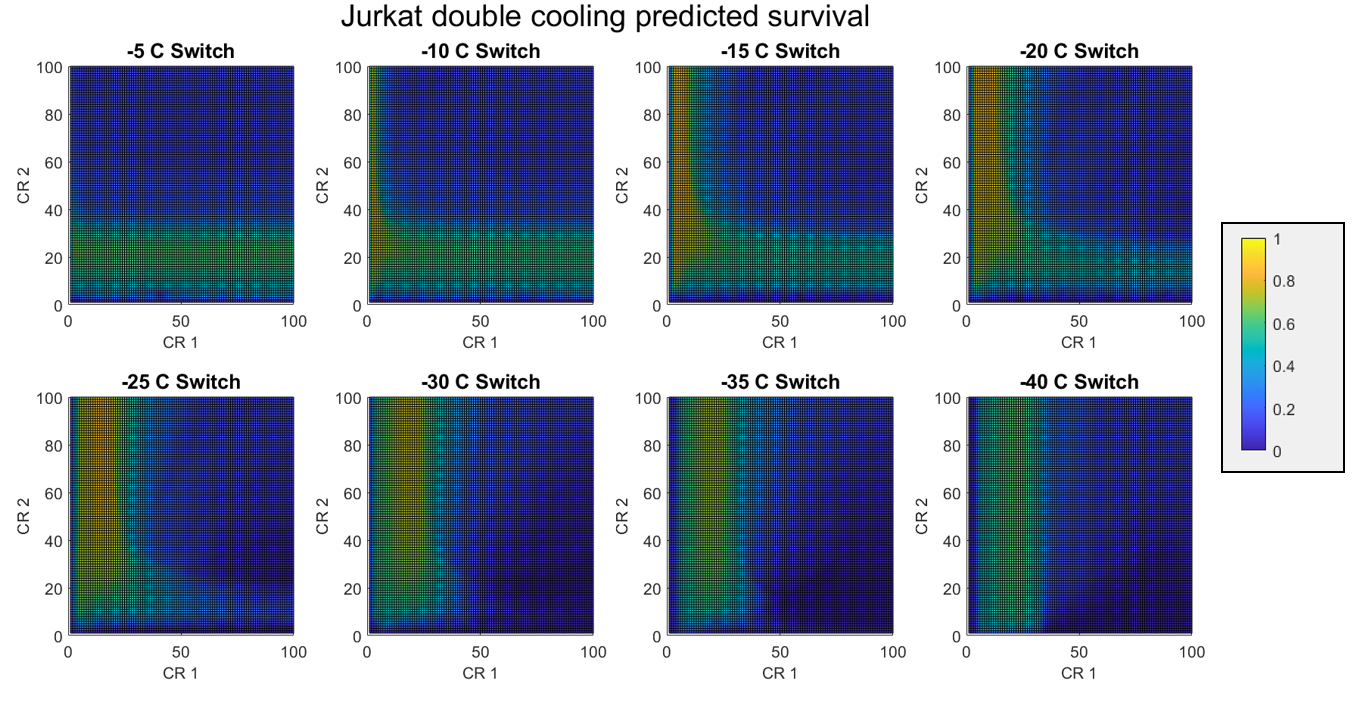}
    \caption[Error corrected Jurkat survival rates]{ Predicted survival rates for Jurkats undergoing two-step cooling with switching temperatures between -5 to -40 \degree C  and cooling rates between 0.1 to 10\degree C $min ^{-1}$. The survival landscapes are obtained by collating all of the data for each switching temperature. The colouring of the survival landscape ranges from blue (Low) to yellow (High).}
    \label{Jurkat Bar graph}
\end{sidewaysfigure}

 A unique relationship was found between CR1, CR2 and the switching temperature(s) during these searches for two-stepped cooling. Similar to standard linear cooling, we found that a maximum post-thaw survival of the cells exists at an absolute global peak. However, unlike single step cooling, multi-step cooling has many approaches to this peak. This data can easily be summarised within a "Survival landscape" for two-step cooling, as is shown in Figure \ref{Jurkat Bar graph}. For three-step cooling this phenomena is not as easy to visualise due to increasing dimensionality. However, we again found that there was a single global optimum for post-thaw survival across available cooling profiles. Unlike single and two-step cooling, there is not necessarily a single peak; instead we observe multiple local peaks but a single global optimum. 
 
\begin{figure}[h]
	\centering
	\includegraphics[scale=0.7]{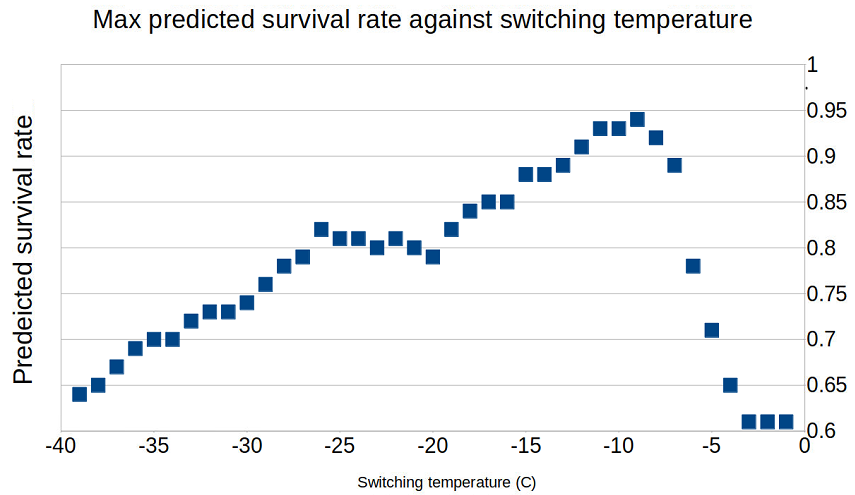}
	\caption[Error corrected Jurkat survival rates curve]{ Maximum Predicted survival rates for Jurkats undergoing two-step cooling. Utilising the optimum predicted cooling trajectory for each switching temperature between 0 to -40 \degree C  and cooling rates between 0.1 to 10\degree C $min ^{-1}$. A maximum predicted post-thaw survival of 0.94 is achieved using a cooling profile of CR1 = 0.1\degree C $min ^{-1}$, CR2 = 9.70\degree C $min ^{-1}$ and Switch  = -9\degree C. Unlike the single peak of standard cooling, we can see a second local optimum at a switching temperature of -25\degree C.}
	\label{Two-stepped all survivals}
\end{figure}

 An absolute global optimum predicted survival of 0.94 with cooling regime CR1 = 0.1 \degree C$min^{-1}$, CR2 = 9.7 \degree C$min^{-1}$ and a switch at -9\degree C. This is displayed in Figure \ref{Two-stepped all survivals}. Unlike standard cooling, it is notable that there is not a single standard upside down "U" curve. Instead we observe a lower survival peak at a switching temperature of -25\degree C in addition to the global optimum at -9\degree C.  

\begin{figure}[h]
	\centering
	\includegraphics[scale=1.2]{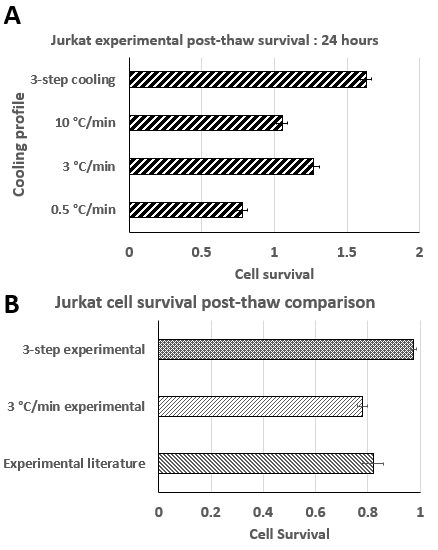}
	\caption[Multi-step vs Standard cooling]{Post-thaw survival of Jurkat cells utilising a 0.5, 3, 10 $\degree C min^{-1}$ and three-step cooling profile (A) 24 hours after thawing. Experimental Jurkat cell survival (B) for 3-step cooling, standard cooling (3\degree $C min^{-1}$) and survival rates achieved by others using standard cooling \cite{YU20172653}. Suppl. Fig. S1 shows viability 1 hour after cooling.}
	\label{Jurkat bar graphs A and B}
\end{figure}

 Finally, we investigated three-step cooling. We achieved a maximum post-thaw survival at 0.98$\pm$0.01 with a cooling profile of CR1 = 1\degree C$min^{-1}$, Switch1 = -13\degree, CR2 = 2\degree C$min^{-1}$, Switch2 = -26\degree, CR3 = 10\degree C$min^{-1}$. This protocol greatly outperforms the linear cooling simulation's maximum predicted survival of 0.72$\pm$0.02. Most importantly, three-step cooling also outperformed the maximum experimental result of 0.8$\pm$0.04.  A comparison between our standard linear model, multi-stepped model and state-of-the-art storage for Jurkat cells can be found in Figure \ref{Jurkat bar graphs A and B}B. 

\begin{figure}[h]
	\centering
	\includegraphics[scale=1.0]{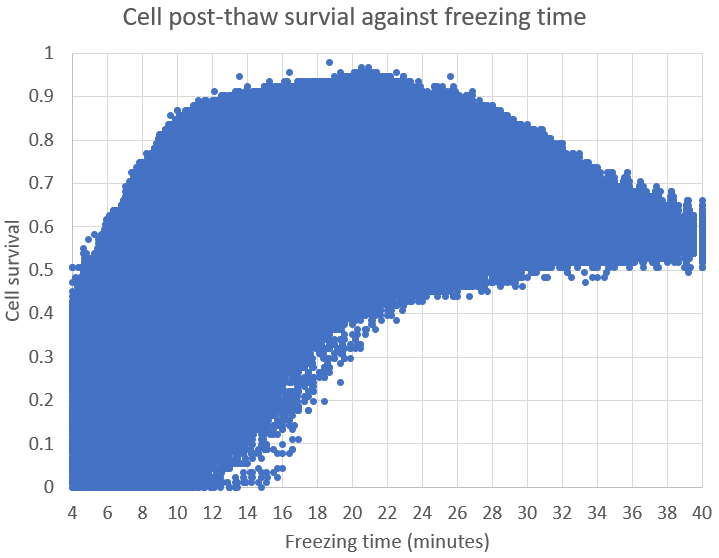}
	\caption[Simulated three stepped cooling time of Jurkat cells with survival rates]{Simulated post-thaw survival of Jurkat cells cryopreserved using 3-step cooling profiles and against freezing time. Notably there is still a clear structure to the overall survival curve and still presents an overall "inverted-U" curve".}
	\label{Cell time}
\end{figure} 

\FloatBarrier
\section{Discussion}

   Using our computational framework we have investigated the post-thaw survival of several cell types undergoing standard cooling profiles. The cell types chosen for study were Jurkat cells, human induced pluripotent stem cells (hiPSCs) and HeLa cells. In addition, the impact of multi-step cooling on the post-thaw survival of cryopreserved Jurkat cells has also been studied. For Jurkat cells this led to a post-thaw survival above that achieved by equivalent cutting edge techniques. 
  
\subsection{Freeze thaw experiments and computational predictions} 

  From our simulations, it was observed that cells cooled slower than the optimum experienced a greater risk of cell death due to an inability to dilute surrounding solutes, leading to a toxic build up of chemicals at the membrane. Furthermore, cells cooled at rates higher than the optimum experienced a decrease in post thaw survival due to an increasing likelihood of IIF. Overall, our results for standard cooling profiles are well matched to the "Inverted-U" survival curve defined by the two-factor hypothesis of Mazur \textit{et al} and others \cite{Whittingham411,mazur1960,mazur1984freezing}.
  
  The results for post-thaw survival results of our computational model are compared against experimental data from multiple sources \cite{YU20172653, STEM, mcgrath1975experimental} yielding excellent agreement (Figure \ref{Simulated cell survivals}). In addition, we have also made comparisons against our own experimental data for Jurkat cells, again yielding excellent agreement (Figure \ref{Jurkat bar graphs A and B}B).
  
  Our computational model for hiPSCs predicted that a maximum post-thaw survival of 0.91 $\pm$0.2 could be achieved using a cooling rate of 1 \degree C $min^{-1}$. This is well matched to Hayashi\textit{et al} who achieved a maximum survival of 0.96 $\pm$0.03 using the same cooling rate \cite{STEM}. Notably, our simulation also displayed the same rapid decrease in cellular post-thaw survival at cooling rates higher than 7\degree C $min^{-1}$ as found in Hayashi \textit{et al}.'s work. This rapid reduction in post-thaw survival of hiPSCs is due to the cell type having greater vulnerability to IIF compared to other cell types \cite{hunt2011cryopreservation}. This accordance shows that our computational framework also predicts cellular responses outside of the optimum. For HeLa cells, our framework predicted a maximum post-thaw survival of 0.43$\pm$0.05 when cooled at 40\degree C $min ^{-1}$. This is slightly higher than the 0.3 achieved by McGrath \textit{et al} using a cooling rate of 65\degree C $min ^{-1}$\cite{MAZURYEAST}. Finally, our framework predicts that a maximum post-thaw survival of 0.71 $\pm$0.02 can be achieved for Jurkat cells using a 3 \degree C $min ^{-1}$ cooling rate. This prediction is well-matched to the experimentally observed optimum of 0.8 $\pm$0.03 at the same cooling rate in G. Yu \textit{et al}'s work \cite{YU20172653}. Overall, across all three cell types our computational framework shows well-matched results to the experimental data for both optimum cooling position and shape of the "Inverted-U" survival curve. 
  
  \subsection{Optimising experimental protocols} 
  
  Another major part of our work was the investigation of the impact of multi-step cooling on the post-thaw survival of cells. To this end we have simulated Jurkat cells undergoing three cooling rate profile types: standard single, two-step and three-step cooling. The objective of this work being to investigate if alternative, more complex cooling profiles could achieve post-thaw survival rates greater than that of current cutting edge techniques. Cooling profiles were simulated within the range of 0 to -40 \degree C utilising cooling rates ranging from 1 to 10 \degree C $min^{-1}$. Jurkat cells were chosen as our target cell type due to their consistent use in multiple areas of biomedical and pharmaceutical research. This includes cell toxicity tests \cite{buenz2004cordyceps,werner2013use} often used in pharmaceutical research and immune cell signal processing \cite{abraham2004jurkat,gaud2018regulatory} for cancer research.  

  An exhaustive search across all simulated protocols allowed us to find the maximum predicted post-thaw survival for one-, two- and three-step cooling. 
  
  Simulations achieved a maximum predicted post-thaw survival of 0.71 $\pm$0.02 for Jurkat cells using a standard cooling profile of 3\degree C $min^{-1}$. This result is well matched to experimental data of others \cite{YU20172653} and our own result of 0.76$\pm$0.06 at the same cooling rate. The simulated and experimental results can be found within Figure \ref{Simulated cell survivals} and \ref{Jurkat bar graphs A and B}B. Two step cooling profiles were predicted to significantly outperform single stepped profiles, achieving a maximum post-thaw survival of 0.94$\pm$0.03 using a cooling profile of CR1 = 0.1\degree C $min^{-1}$, CR2 = 9.70\degree C $min^{-1}$ and a switching temperature of -9\degree C. Notably, more than 13 of the possible 40 switching temperatures outperformed standard single step cooling. A summary of the maximum post-thaw survival achieved by each switching temperature can be found in Figure \ref{Two-stepped all survivals}. Finally, three stepped cooling simulations predicted a maximum post-thaw survival of 0.98$\pm$0.03 could be achieved with  CR1 = 1\degree C $min^{-1}$, T$_{Switch1}$ = -13\degree C, CR2 = 2\degree C $min^{-1}$, T$_{Switch2}$ = -26\degree C , CR3 = 10\degree C $min^{-1}$. Overall, our simulation results indicated that three-stepped cooling profiles can significantly increase the post-thaw survival of cryopreserved Jurkat cells.   
  
  To validate our computational results we conducted wet-lab experiments for Jurkat cells using one-stepped cooling rates of 0.5, 3 and 10\degree C $min^{-1}$ and our computationally derived three-stepped profile. Post-thaw cell survival after three-stepped Jurkat cell cooling significantly exceed single step cooling results. In particular, we achieved a post-thaw survival of 0.98$\pm$0.03 after 1 hour post-thaw. These results are displayed in supplementary Figure 1. Post-thaw survival after 24 hours was also assessed by comparing thawed population counts against pre-freeze values. The 3 and 10\degree C $min^{-1}$ populations had slightly more cells than pre-freeze with the 0.5 \degree C $min^{-1}$ population having a smaller population than pre-freeze numbers. Overall, these results demonstrate the ability of the three-stepped cooling profile to significantly improve post-thaw survival of Jurkat cells.
  
  Several other works have also studied cryopreservation protocols using a combination of computational and experimental techniques. For instance, Pollock \textit{et al} \cite{pollock2017algorithm} utilised a convergence algorithm to optimise the freeze/thaw of Jurkat cells, achieving a maximum post-thaw survival of 0.90$\pm$0.02. However, Pollock's work has several major differences to our own. Firstly, the approach of Pollock \textit{et al} requires a high-throughput experimental approach to obtain large numbers of samples based on various CPA types, concentrations and cooling rates. Additionally, their work requires continual iterative feedback from experimental work until an optimum is found. The requirement of repeat experimental verification and updating of results means a much more time consuming process. Alternatively, the experimentally focused liquidus tracking (LT) approach was used by Kay\textit{et al}\cite{kay2015liquidus} and Puschmann \textit{et al} \cite{puschmann2017liquidus} to optimise CPA concentrations and cooling rates. Unlike Pollock\textit{et al}'s work, this methodology relies on the real time alteration of CPA concentrations via external pumps added to the freezer. Using this method Puschman managed to increase the post-thaw survival of cryopreserved alginate encapsulated liver cells (AELCs) from 60\% to 90\%. This increase in survival is comparable to that achieved by our own work when utilising multi-stepped cooling profiles. In addition to the increased experimental demands, a downside of the LT method is the need to optimise the CPA mix before the LT process itself can be used. Thus it requires significant trial and error wet lab work before optimisation can begin. 
  
  The hybrid mathematical optimisation approach of Y. Hayashi \textit{et al} \cite{STEM} is the most comparable to our own work. The authors also made predictions for cells cryopreserved using multi-stepped cooling profiles. Notably, they made several observations that differ from the findings of our own work. Firstly, they predicted that a considerable number of cooling profiles can achieve the same high post-thaw survival as the optimum. This is contrary to our results which indicate that there is an overall increase to an optimum value, as shown in Figure \ref{Cell time}. This discrepancy is likely due to the lack of CPA's in Hayashi's simulations, which may lead to different dynamics due to the different extracellular conditions. However, for slow cooling we believe that the inclusion of CPA's is paramount due to their prevalence in the industry and the significant impact they have on cellular post-thaw survival.
  
  Notably, the optimum cooling rates presented by Hayashi utilise a single cooling rate from 0 to -40/50$\degree $C at which point they switch to a slower cooling rate until reaching -60$\degree$C. This significantly differs from our work, as we found that utilizing multiple cooling rates before reaching -40$\degree$C produced the greatest post-thaw survival. Additionally, the authors' top profiles often followed a pattern of $C2 < C1$  and  $C3 > C1$, which is contrary to our own finding that optimum cooling profiles followed the form of $C1 < C2 < C3$. A possible reason for this contradiction is the use of different cell types in our and Hayashi's study. However, this cannot be confirmed due to Hayashi's findings, as of yet, not being experimentally validated. 
  
\subsection{Limitations} 

 Our work is limited by its reliance on exact experimental protocols. Our computational framework thus only predicts the response of the specified cell type with regards to exact CPA, salt and extracellular medium concentrations. Thus, if any change is made to an initial experimental setup; all simulations must be ran again to accurately predict a cells response to other parameters. Thus, in future we will expand our model to be more flexible with predictions that are made from initial conditions.
 
 Our computational approach uses a relatively simplistic grid search method for optimising these multi-stepped profiles. In future, we intend to improve our computational frameworks by using more complex optimisation techniques. For example, the use of a closed loop variable optimisation algorithm would likely significantly reduce both the time to optimise cooling profile variables and the number of simulations which need to be ran. 
 
 Jurkat cells were chosen for the multi-stepped cooling investigation over other cell types due to experimental availability and cooling rate limitations of available freezers. In future we hope to extend our work by testing multiple cell types across a greater range of cooling rates.
 
 As we only utilised a maximum of three cooling rates in our multi-stepped cooling protocol, further work is required to investigate more complex cooling profiles. Ultimately, we hypothesise that an accelerating cooling rate may prove optimum for cellular post-thaw survival. Finally, we have not tested the efficiency of our model for large-scale cryo-banking (Volume $>$ 200 mL).

 Further improvements for predicting ice formation and crystal growth could be made via the addition of ice crystallisation modelling \cite{boutron1986comparison} to the currently existing one. This would allow for improvements to the kinetics.

\subsection{Conclusion} 

 Here, we presented two major results. Firstly, a general computational framework can be utilised to mechanistically model and accurately predict the post-thaw survival of cells in suspension undergoing cryopreservation. 
 
 Secondly, we find that a multi-stepped cooling profile correctly predicts for an up to one order of magnitude reduction in cell death. This prediction was validated experimentally. Cryopreserved Jurkat cells were found to have a order of magnitude reduction in cell death post-thaw when utilising a multi-stepped cooling rate. 
  
 Overall, within this work we present novel computational modelling work for improving \textit{in vitro} cryopreservation.  

\section{Declarations}

\subsection{Author contributions}

Jack Jennings: Executed the project, formal analysis, investigation and methodology, writing and editing of the manuscript and co-designed the study.
\\
Sanja Bojic: Performed experimental work, performed wet lab procedures, maintained cells, key in developing experimental methodology, assisted in research direction and reviewing of manuscript.  
\\
Lukas Breitwieser: Assisted in computational design and testing, helped with reviewing the manuscript. 
\\
Alex Sharpe: Assisted in experimental work and wet lab procedures. 
\\
Roman Bauer: Acquired financial support for the study, co-designed the study, assisted in research and development of the
work as well as reviewing the manuscript

\subsection{Funding}

This work was supported by the Engineering and Physical Sciences Research Council of the UK (EP/S001433/1, EP/S001433/2) (RB) and via a PhD studentship funded by Newcastle University’s School of Computing (Newcastle upon Tyne, UK) (JJ). LB was supported by the CERN Knowledge Transfer office.

\subsection{Acknowledgements}

We would like to acknowledge the BioDynaMo team and consortium (www.biodynamo.org) for their assistance across the development of this project.

\subsection{Conflicts of interest}
The authors declare that they have no conflicts of interest.

\bibliographystyle{plainnat}
\bibliography{sample}

\begin{figure}[h]
	\centering
	\includegraphics[scale=1.0]{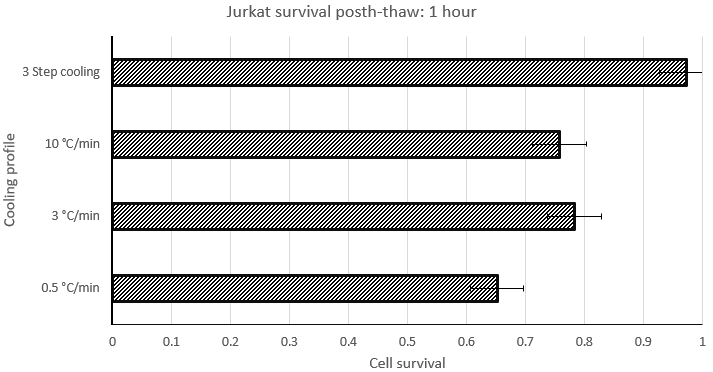}
	Supplementary Figure S1. Multi-stepped vs Standard cooling. Post-thaw survival of Jurkat cells utilising a 0.5, 3, 10 \degree C\,min$^{-1}$ and 3 step cooling profile. Survival 1 hour after thawing is shown.
	\label{fig:SuppFig1}
\end{figure} 

\begin{xltabular}{\linewidth}{ l | X }
  \caption*{Supplementary Table S1. Information on variables used in this study.} 
 \label{table:SuppTab1}\\
 \hline \hline

\textbf{\normalsize Variables} & \textbf{\normalsize Definition and/or relevant information}  \\
 \hline 
\endfirsthead
 \hline \hline

\endhead
\textbf{Effective surface area} &$A$ [$m^2$] \\ \hline
\textbf{Diffusion coefficient} &$\alpha$ \\ \hline
\textbf{Cooling rate} &$B$ [$K s{-1}$] \\ \hline
\textbf{Activation energy} &$E_a$ [$J mol^{-1}$] \\ \hline
\textbf{Latent heat of fusion} &$\Delta H_f$ [$J kg^{-1}$] \\ \hline
\textbf{Water Permeability}  &$L_p$ [$m s^{-1} Pa^ {-1}$] \\ \hline
\textbf{Molar amount} &$n$ [$mol$] \\ \hline
\textbf{CPA permeability} &$p_s$ [$m s^{-1} Pa^{-1}$] \\ \hline
\textbf{Probability of intracellular ice formation} &$P_{IIF}$\\ \hline
\textbf{Probability of solution effect} &$P_{SOL}$\\ \hline
\textbf{Probability of survival} &$P_{SUR}$\\ \hline
\textbf{Courant-Friedrichs-Lewy (CFL) number} &$Q$ [$s$], defined in Equations \ref{CFL} and \ref{CFL 3D} \\ \hline
\textbf{Universal gas constant} &$R$ [$J mol^{-1} K ^{-1}$] \\ \hline
\textbf{Time} &$t$ [$s$] \\ \hline
\textbf{Temperature} & T [$K$] \\ \hline 
\textbf{Concentration} &$u$ [$mol m^{-3}$] \\ \hline
\textbf{Cellular water volume} & V [$m^3$]\\ \hline 
\textbf{Partial molar volume} &$v$ [$K s{-1}$] \\ \hline
\textbf{Spatial variables} & x, y and z [$m$] \\ \hline

 \textbf{Reference} & 0  \\ \hline 
 \textbf{Solute} & s  \\ \hline
 \textbf{Water} & w  \\ \hline
 \textbf{Spatial position} & i =x , y=j and k=z  \\ \hline
 \textbf{Switching position} & switch  \\ \hline
 \textbf{Intracellular ice formation} & IIF  \\ \hline
 \textbf{Surface catalysed nucleation} & SCN  \\ \hline
 \textbf{Volume catalysed nucleation} & VCN  \\ \hline
 \textbf{Osmotically non-interacting} & b  \\ \hline
 \textbf{Position in time } & n  \\ \hline
\textbf{Surface catalysed nucleation} & SCN  \\ \hline
\textbf{Volume catalysed nucleation} & VCN  \\ \hline

\end{xltabular}

\end{document}